\documentstyle[12pt]{article}
\title{
{\vspace{-3cm} \normalsize \hfill MS-TPI-92-26
                                            }\\[25mm]
Study of the complex fermion determinant in a
$\rm U(1)_L \otimes U(1)_R$ symmetric Yukawa model}

\author{Gernot M\"unster and Markus Plagge \\
        Institut f\"ur Theoretische Physik I,
        Universit\"at M\"unster\\
        Wilhelm-Klemm-Str.~9, D-4400 M\"unster, Germany}
\date{November 11, 1992}

\newcommand{\be}{\begin{equation}}
\newcommand{\ee}{\end{equation}}
\newcommand{\ul}[1]{\underline{#1}}

\begin{document}
\maketitle

\begin{abstract}
Lattice theories that contain chiral multiplets of fermions can have
complex fermion determinants.
This is for example the case for the $\rm U(1)_L \otimes U(1)_R$
symmetric Yukawa model with mirror fermions, if the number of
generations of fermions and mirror fermions is odd.
Whether a numerical simulation of such a model is possible depends on
the magnitude of fluctuations of the complex phase factor of the fermion
determinant.
We investigate the fermion determinant of the U(1) Yukawa model with
mirror fermions for a physically relevant choice of parameters.
The argument of the complex phase turns out to fluctuate only very
little and is at most of the order of $2 \cdot 10^{-3}$.
\end{abstract}
%
\section{Introduction}
The problem of formulating theories with a chiral fermion content on a
lattice has not yet been solved in a satisfactory way.
As is well known lattice fermions are accompanied by unwanted doublers
which one likes to remove at least in the continuum limit.
Their appearance is related to the Nielsen-Ninomiya theorem
\cite{NiNi}, which requires that to every lattice fermion with a given
set of quantum numbers a mirror fermion with the same quantum numbers
but opposite parity exists.
For a naive discretisation of the fermion action half of the doublers
play the role of mirror fermions \cite{KarSmit}.

The Wilson term \cite{Wilson}, which achieves the removal of doublers in
vectorlike theories, spoils chiral invariance and cannot be used
directly in chiral theories.
For fermions interacting with scalar fields via Yukawa interactions a
chiral invariant Wilson-Yukawa term has been proposed \cite{SmitSwift},
but there is evidence that it does not fulfill its goal \cite{Bock}.
On the other hand it is possible to remove the doublers in chiral Yukawa
models with the help of a Wilson-like term but at the cost of
introducing extra mirror fermions \cite{MirFer}.
This yields the minimal possible proliferation of extra fermions.
If one does not like to have the mirror fermions in the physical
spectrum, even at a higher mass scale, the task is to find out whether
the mirror fermions can be removed or decoupled in the continuum limit.

If gauge fields are neglected, chiral Yukawa models with mirror fermions
are for a particular choice of parameters invariant under a
Golterman-Petcher shift symmetry \cite{GolPet} with respect to the
mirror fermion field \cite{LinWit}.
As a result the mirror fermions decouple from the fermions and from the
scalar field.
It remains a truly chiral set of fermions interacting with the scalar
field.

For a numerical simulation of such theories with the Hybrid Monte
Carlo algorithm, however, the fermion fields have to be duplicated
again.
This is due to the following fact.
The part of the action bilinear in the fermion fields can be written
with the help of the ``fermion matrix'' $Q(\phi)_{yx}$ as
\be
S_f = \sum_{x,y} \bar{\Psi}_x Q(\phi)_{xy} \Psi_y \,,
\ee
where
\be
\Psi_x = \left( \begin{array}{c}
                \psi_x \\ \chi_x
                \end{array} \right)
\ee
contains fermion fields $\psi$ and mirror fermion fields $\chi$,
and $\phi$ is the complex scalar field.
The matrix $Q$ is not necessary positive definite, while the hybrid
Monte Carlo algorithm requires a positive definite fermion matrix.
A duplication of fermion fields by introducing two generations
$\Psi^{(1)}$ and $\Psi^{(2)}$ with opposite chirality amounts to
replacing $Q$ by $Q^{+}Q$ and ensures the positive definiteness.
This duplication is of a purely algorithmic origin and has nothing to
do with the fermion doubling problem.
Therefore it is desirable to find ways to simulate these models without
additional duplication.

In particular, for a $\rm U(1)_L \otimes U(1)_R$ symmetric model the
determinant of $Q$ is in general complex, preventing a simulation of
this model with standard Monte Carlo algorithms.
Nevertheless a simulation without duplication would be possible
depending on the fluctuations of the phase angle $\alpha$ of the fermion
determinant.
Two possible situations can be imagined:
\begin{enumerate}
\item
the angle $\alpha$ fluctuates only very little about zero.
Then a simulation with $| \det Q |$ appears feasible and the phase
factor $\exp i \alpha$ can be put into the observables or neglected
completely.
\item
$\alpha$ fluctuates strongly and a simulation without taking the phase
of $\det Q$ into account is not possible.
\end{enumerate}
In order to find out which of these two possibilities holds we have
investigated the fermion determinant in the $\rm U(1)_L \otimes U(1)_R$
symmetric Yukawa model with mirror fermions.
\section{Calculation of the fermion determinant}
The $\rm U(1)_L \otimes U(1)_R$ symmetric Yukawa model with mirror
fermions has been considered in ref.\,\cite{U1sym} in detail and we
refer to this article for the definition of the action and the
parameters.
The $8 \otimes 8$ matrix $Q$ is given in a $2 \otimes 2$ block notation
by
$$
Q(\phi)_{xy} = \delta_{xy} \left(
\begin{array}{cccc}
  G_{\psi} \phi_x^+  &  0          &  1          &  0            \\
  0          &  G_{\psi} \phi_x    &  0          &  1            \\
  1          &  0          &  G_{\chi} \phi_x    &  0            \\
  0          &  1          &  0                  &  G_{\chi} \phi_x^+
\end{array}                            \right)
$$
\be
- K \sum_{\mu} \delta_{x,y+\hat{\mu}}   \left(
\begin{array}{cccc}
 0                  & \Sigma_{\mu} & r & 0     \\
 \bar{\Sigma}_{\mu} & 0            & 0 & r     \\
 r & 0 & 0                  & \Sigma_{\mu}     \\
 0 & r & \bar{\Sigma}_{\mu} & 0
\end{array}                            \right) \ .
\ee
Here $x$ and $y$ are lattice points, the sum $\sum_{\mu}$ runs over
eight directions of the neighbours, and $\hat{\mu}$ is the unit vector
in the direction of $\mu$.
The Euclidean $\gamma$-matrices are expressed in a chiral basis as
\be
\gamma_{\mu} = \left(
\begin{array}{cc}
0                  & \Sigma_{\mu}  \\
\bar{\Sigma}_{\mu} & 0
\end{array}                       \right) ,
\hspace{1cm}
\gamma_5 = \left(
\begin{array}{cc}
  1  &  0  \\
  0  & -1
\end{array}       \right) \,.
\end{equation}
For $\mu=1,2,3$ we have
$\Sigma_{\mu} = -\bar{\Sigma}_{\mu} = -i\sigma_{\mu}$ and
$\Sigma_4 = \bar{\Sigma}_4 = 1$, where
$\sigma_{1,2,3}$ denote the Pauli-matrices.
For negative indices the definition is given by
$\Sigma_{\mu} \equiv -\Sigma_{-\mu}$.
The Yukawa couplings of the fermions and mirror fermions are $G_{\psi}$
and $G_{\chi}$, respectively.
The term proportional to $K r$ is a chiral invariant Wilson-term
which serves to give the fermion doublers masses of the order of the
cutoff.
The Wilson parameter $r$ is set equal to 1 usually.
Finally $K$ is the fermionic hopping parameter, whose critical value is
1/8 for vanishing Yukawa couplings.

The hybrid Monte Carlo algorithm requires the inversion of $Q$, which
is done by conjugate gradient or minimal residue algorithm.
A calculation of the determinant of $Q$ is more demanding.
Straightforward application of standard determinant algorithms is not
possible owing to the need of large storage.
We proceeded in the following way.
Let $L_i,\ \ i=1,2,3,4$ be the size of the lattice in the direction $i$.
We consider even $L_i$ and set $L_4 = 2n$.
$Q$ can be considered as a $2n \times 2n$ matrix consisting of
block matrices belonging to time slices.
Each block itself is a $M \times M$ matrix with $M = 8 L_1 L_2 L_3$.
In a first step the matrix is reordered by an even number of exchanges
of rows and columns such that a band structure is obtained:
\be
\left( \begin{array}{cccccccccc}
 m_{T(1)} & b        & a        &          &       &       & & & & \\
        a & m_{T(2)} & 0        & b        &       &       & & & & \\
        b & 0        & m_{T(3)} & 0        & a     &       & & & & \\
          & a        & 0        & m_{T(4)} & 0     & b     & & & & \\
          &          & b        & 0        & \cdot &       & & & & \\
          &          &          &          &       & \cdot & & & & \\
  & & & & & & \cdot &       &             &          \\
  & & & & & &       & \cdot & 0           & \pm b    \\
  & & & & & &       & 0     & m_{T(2n-1)} & a        \\
  & & & & & &       & \pm a & b           & m_{T(2n)}
            \end{array} \right).
\ee
The blocks $a$ and $b$ are constant, whereas the $m_j$ depend on the
scalar field $\phi$.
The indicated signs near the lower right corner depend on the temporal
boundary conditions.
$T(j)$ is a permutation of the time slices defined by
$$
T(1) = n,\ \ T(2n) = 2n,
$$
\be
 T(2j) = n-j,\ \ T(2j+1) = n+j \ \ \
\mbox{for}\ j = 1,\ldots,n-1 \,.
\ee
The matrix above is now dealt with by LU-block factorisation
\cite{GoLo}.
The resulting upper triangular matrix has blocks $D_k$ on its diagonal,
which can be determined recursively from the blocks $m_j , a$ and $b$.
The complete determinant is given by
\be
\det Q = \prod_{k=1}^{2n} \det D_k \,.
\ee
The ``little'' matrices $D_k$ are not sparse.
But for a $4^3 \cdot L_4$ lattice they are small enough such that their
determinants can be calculated by standard numerical routines.
Similar techniques have been used by Toussaint as indicated in
\cite{Toussaint}.

The storage needed does not depend any longer on $L_4$ and amounts to
$6144 (L_1 L_2 L_3)^2$ bytes using complex extended precision.
In our case this is slightly more than 3 Megawords, which means a
reduction by a factor of about 1/10 compared to a standard treatment.
The CPU time is proportional to $L_4$ instead of $L_4^3$, which
standard routines require.
For $L_4 = 8$ it takes 3 minutes to calculate one determinant on the
Cray Y-MP with a high degree of vectorisation (300 MFlops).

Because the correctness and precision of the algorithm is crucial we
have tested it in different ways:
\begin{enumerate}
\item For lattices smaller than $4^4$ a comparison with the results
obtained with the help of IMSL-library routines has been made.
\item For constant scalar fields $\phi$ the determinant can be
calculated exactly by Fourier transformation, and we compared with the
resulting values.
\item The relation
\be
\frac{d}{d \lambda}
[ \ln \det (Q+\lambda I_{uv}) ]_{\lambda=0}
 = Q^{-1}_{vu}
\ee
with $(I_{uv})_{xy} = \delta_{ux} \delta_{vy}$ and variable scalar field
was verified numerically, where the left hand side was evaluated by our
determinant algorithm and the right hand side with the help of the
conjugate gradient algorithm.
\item For particular choices of the Yukawa couplings the determinant has
to be real.
We checked this property to a high precision.
\end{enumerate}
In all cases the deviations were of the size of the numerical precision
of the computer.
For example in the last mentioned item the ratio of the imaginary to
the real part of the determinant was of the order of $10^{-14}$.
\section{Results on the fermion determinant}
Some general results on the fermion determinant in the
$\rm U(1)_L \otimes U(1)_R$ symmetric Yukawa model are available.
\begin{enumerate}
\item The expectation value $\overline{\det{Q}}$ is real.
This is due to
$\det Q(\phi^+) = (\det Q(\phi))^*$,
and the weight of a scalar field $\phi$ in the functional integral being
the same as that of its complex conjugate $\phi^+$.
\item For $G_{\psi} = \pm G_{\chi}$ the determinant is real for each
single scalar field configuration, as can be shown using the symmetries
of the action.
\end{enumerate}
For other physically interesting choices of parameters the aim is to
obtain information about the size of the fluctuations of the phase angle
$\alpha$ defined by
\be
\det Q = |\det Q| \, e^{i \alpha} \,.
\ee

We have investigated the fermion determinant in four series of points in
the parameter space.
Each series is specified by the values of $G_{\psi}$, $G_{\chi}$, $K$
and the quartic scalar self-coupling $\lambda$.
The parameters are summarized in table 1.
Series D has a real determinant and serves as a check on the program.

Within each series the scalar hopping parameter $\kappa$ was varied in
such a way that five equidistant points in parameter space are obtained,
which start in the symmetric phase near a scalar mass of $m_R \approx 1$
and end in the phase with broken symmetry at a scalar mass of the same
magnitude.
This covers the physically interesting region near the second order
phase transition line.
At each of the 20 points in parameter space after equilibration the
determinant was calculated 40 times, always separated by hundred
trajectories from each other.
The results of the calculation are displayed in table 2.
It displays the average values of the real and the imaginary parts of
$\det Q$, together with the average value $\overline{\alpha}$ of the
phase and its standard deviation
\be
\Delta \alpha = \frac{1}{40} \sum_{i=1}^{40}
(\alpha_i - \overline{\alpha})^2 \,.
\ee
In series D, where the phase is known to be zero, the standard deviation
of $\alpha$ yields an estimate of the precision of the numerical
results.
\begin{table}[tb]
\begin{center}     \Large\bf Table 1  \rm\normalsize
\end{center}
Parameters of the four series of points at which the fermion determinant
has been measured.
\begin{center}
\begin{tabular}{|c|c|c|r|c|} \hline
label & $\lambda$ & $G_{\psi}$ & $G_{\chi}$ & $K$ \\ \hline
A     & 1.0       & 1.0        &   0.0      & 0.10 \\
B     & 1.0       & 0.3        &   0.0      & 0.10 \\
C     & $10^{-4}$ & 0.1        & - 0.2      & 0.13 \\
D     & 1.0       & 0.3        & - 0.3      & 0.10 \\ \hline
\end{tabular}
\end{center}
\vspace{1cm}
\begin{center}     \Large\bf Table 2  \rm\normalsize
\end{center}
Results of the numerical calculation of the complex fermion determinant.
\begin{center}
\begin{tabular}{|c|c||l|r@{$\,\cdot\,$}l|r@{$\,\cdot\,$}l|l|}
\hline
label & $\kappa$ & \rule{0pt}{14pt} $\overline{\mbox{Re\,} \det{Q}}$ &
\multicolumn{2}{c|}{\rule{0pt}{14pt} $\overline{\mbox{Im\,} \det{Q}}$} &
\multicolumn{2}{c|}{\rule{0pt}{14pt} $\overline{\alpha}$} &
\hspace{5mm} $\Delta \alpha$ \\ \hline
\rule{0pt}{14pt}
  & $0.070$ & $6.90 \cdot 10^{7}$  & $2.68$  & $10^{4}$  & $3.89$ &
  $10^{-4}$ & $2.04 \cdot 10^{-3}$ \\
  & $0.079$ & $2.20 \cdot 10^{8}$  & $-2.48$ & $10^{5}$  & $2.54$ &
  $10^{-5}$ & $2.67 \cdot 10^{-3}$ \\
A & $0.088$ & $3.34 \cdot 10^{11}$ & $7.07$  & $10^{7}$  & $-6.17$ &
  $10^{-4}$ & $2.23 \cdot 10^{-3}$ \\
  & $0.097$ & $9.65 \cdot 10^{13}$ & $-6.46$ & $10^{10}$ & $2.95$ &
  $10^{-4}$ & $2.16 \cdot 10^{-3}$ \\
  & $0.105$ & $7.70 \cdot 10^{18}$ & $-2.14$ & $10^{15}$ & $-4.11$ &
  $10^{-4}$ & $2.16 \cdot 10^{-3}$ \\ \hline
\rule{0pt}{14pt}
  & $0.137$ & $1.67 \cdot 10^{3}$  & $4.51$  & $10^{-3}$ & $1.15$ &
  $10^{-6}$ & $1.90 \cdot 10^{-5}$ \\
  & $0.145$ & $2.03 \cdot 10^{3}$  & $5.28$  & $10^{-4}$ & $5.55$ &
  $10^{-7}$ & $2.24 \cdot 10^{-5}$ \\
B & $0.153$ & $2.64 \cdot 10^{3}$  & $-1.35$ & $10^{-2}$ & $-3.55$ &
  $10^{-6}$ & $2.18 \cdot 10^{-5}$ \\
  & $0.160$ & $3.09 \cdot 10^{3}$  & $-2.10$ & $10^{-2}$ & $-1.61$ &
  $10^{-6}$ & $2.52 \cdot 10^{-5}$ \\
  & $0.168$ & $5.46 \cdot 10^{3}$  & $-1.98$ & $10^{-2}$ & $-2.34$ &
  $10^{-6}$ & $2.14 \cdot 10^{-5}$ \\ \hline
\rule{0pt}{14pt}
  & $0.077$ & $5.72 \cdot 10^{35}$ & $-5.38$ & $10^{31}$ & $1.54$ &
  $10^{-5}$ & $1.70 \cdot 10^{-4}$ \\
  & $0.081$ & $5.77 \cdot 10^{36}$ & $-1.81$ & $10^{33}$ & $-5.58$ &
  $10^{-5}$ & $2.52 \cdot 10^{-4}$ \\
C & $0.086$ & $9.24 \cdot 10^{38}$ & $4.56$  & $10^{35}$ & $-2.20$ &
  $10^{-5}$ & $2.45 \cdot 10^{-4}$ \\
  & $0.091$ & $3.95 \cdot 10^{42}$ & $-2.30$ & $10^{38}$ & $-1.16$ &
  $10^{-5}$ & $2.57 \cdot 10^{-4}$ \\
  & $0.096$ & $2.20 \cdot 10^{65}$ & $-8.98$ & $10^{60}$ & $3.42$ &
  $10^{-5}$ & $1.53 \cdot 10^{-4}$ \\ \hline
\rule{0pt}{14pt}
  & $0.104$ & $3.90 \cdot 10^{84}$ & $-1.39$ & $10^{70}$ & $1.81$ &
  $10^{-16}$ & $5.50 \cdot 10^{-15}$ \\
  & $0.109$ & $1.10 \cdot 10^{85}$ & $-2.73$ & $10^{70}$ & $-7.16$ &
  $10^{-16}$ & $6.53 \cdot 10^{-15}$ \\
D & $0.114$ & $3.38 \cdot 10^{86}$ & $2.52$  & $10^{72}$ & $9.54$ &
  $10^{-16}$ & $8.06 \cdot 10^{-15}$ \\
  & $0.119$ & $2.55 \cdot 10^{88}$ & $-1.02$ & $10^{74}$ & $-1.68$ &
  $10^{-16}$ & $7.65 \cdot 10^{-15}$ \\
  & $0.124$ & $1.27 \cdot 10^{89}$ & $2.44$  & $10^{75}$ & $2.57$ &
  $10^{-15}$ & $7.42 \cdot 10^{-15}$ \\
\hline
\end{tabular}
\end{center}
\end{table}

{}From the data it appears that nothing particular happens to the
determinant when the vicinity of the phase transition is passed.
The absolute values of both the real and the imaginary parts increase
continuously from the symmetric to the broken phase.
The average values of $\alpha$ are always statistically consistent with
zero, as they should.

Most important is the observation that in all points the standard
deviation of $\alpha$ is very small.
Its value in series A, where it is largest, is near $2 \cdot 10^{-3}$.
In this series the Yukawa couplings differ most in their absolute
values.
This represents the case that is farthest from $G_{\psi} = \pm
G_{\chi}$, where $\alpha$ vanishes.
\section{Conclusion}
The investigation of the complex fermion determinant in the
$\rm U(1)_L \otimes U(1)_R$ symmetric Yukawa model with mirror
fermions has shown that the phase of the determinant is fluctuating only
very little, while the modulus of the determinant varies over many
orders of magnitude in the physically interesting region.
Thus a simulation of the model without further technical doubling of the
number of fermions appears feasible.
This could be achieved by a Hybrid Classical--Langevin algorithm based
on the effective action $S(\phi) - 1/2 \ln \det QQ^+$.

{\bf Acknowledgement}:
We thank I.~Montvay for discussions.
The calculations have been performed on the Cray Y-MP of HLRZ, J\"ulich.

\end{document}